\documentclass[twocolumn,showpacs,preprintnumbers,amsmath,amssymb]{revtex4}
\usepackage{mathrsfs}
\usepackage{amssymb}
\usepackage{graphics}% Include figure files
\usepackage{dcolumn}% Align table columns on decimal point
\usepackage{bm}% bold math
\begin{document}

\title{Guided modes in graphene waveguides}
% title is needed to be considered again

\author{Fan-Ming Zhang$^{1}$}

\author{Ying He$^{1}$}

\author{Xi Chen$^{1,2}$\footnote{Author to whom correspondence should be addressed; Email address: xchen@shu.edu.cn}}

\affiliation{$^{1}$ Department of Physics, Shanghai University,
200444 Shanghai, People's Republic of China}

\affiliation{$^2$ Departamento de Qu\'{\i}mica-F\'{\i}sica, UPV-EHU,
Apdo 644, 48080 Bilbao, Spain}

\date{\today}% It is always \today, today,

\begin{abstract}
By analogy of optical waveguides, we investigate the guided modes in
graphene waveguides, which is made of symmetric quantum well. The
unique properties of the graphene waveguide are discussed based on
the two different dispersion relations, which correspond to
classical motion and Klein tunneling, respectively. It is shown that
the third-order mode is absent in the classical motion, while the
fundamental mode is absent in the Klein tunneling case. We hope
these phenomena can lead to the potential applications in
graphene-based quantum devices.
\end{abstract}

%\keywords{single modes; graphene waveguide; chiral carriers } %Use showkeys class option if keyword
\maketitle

%----------------------------------------------------------------

There has been great interest in the investigations of graphene
\cite{AV0709,C.W.B,M.I.K}, since the two-dimensional form of carbon
named graphene was realized experimentally by A. Geim \textit{et
al}. in 2004 \cite{s306}. This material is densely packed into a
honeycomb structure, which is made out of two distinct triangular
sublattices, is labeled by A and B. The low energy band structure of
graphene is gapless, with massless chiral carriers. These unusual
structure is thus the root of special phenomenon such as anomalous
quantum Hall effect \cite{Nat1,Nat2,v.p.g}, minimum conductivity
\cite{Nat1,Nat2}, and Klein tunneling \cite{C.W.B,Na Ph}. It is also
interesting that the Klein paradox describes a phenomenon that the
relativistic electron can pass through a high barrier perfectly in
contrast to the conventional tunneling. These phenomena are expected
to play an important role in the future nanoelectronic devices.

It is well established that electronic analogies of many optical
behaviors have been achieved in two-dimensional electron gas (2DEG)
system \cite{Datta,Dragoman}. Since the mean-free path approaches to
the order of micron at room temperature, ballistic electrons behave
the quantum wave nature of electron, that is, electrons can refract,
reflect, and interfere in a manner analogous to electromagnetic
waves in dielectrics \cite{T.K1,T.K2,Prl}. Furthermore, electron
waveguides \cite{T.K3,13442,5386,RQY,TNO,OE,CC} have also been
extensively demonstrated theoretically and experimentally. Very
recently, graphene opens a way to study different optics-like
phenomena \cite{pn,nanolett,0808,Beenakker,0806,0804}, such as
negative refraction \cite{pn}, Goos-H\"{a}nchen effect
\cite{Beenakker}, Bragg reflection \cite{0806}, and graphene
waveguide \cite{0804}.

In this Letter, we investigate the guided modes in graphene quantum
well (QW) \cite{PRB74}, acting as a slab waveguide for electron
waves in a form similar to integrated optics. As a matter of fact,
the waveguide configuration of graphene $\mbox{p-n-p}$ junction
considered here can be tailored to desired form and size. For
example, a gate-tunable graphene device can be fabricated from a
single-layer graphene sheet, which is modulated by $n^{++}
\mbox{Si}$ substrate as the back gate and the $\mbox{Ti Au}$ strip
as the top gate. The carrier density in the bulk of the graphene
sheet is tuned by a voltage applied to the back gate, and the
density in the narrow strip below the top gate is tuned by a voltage
applied to the top gate \cite{236803}.

\begin{figure}[]
	\begin{center}
		\scalebox{0.50}[0.50]{\includegraphics{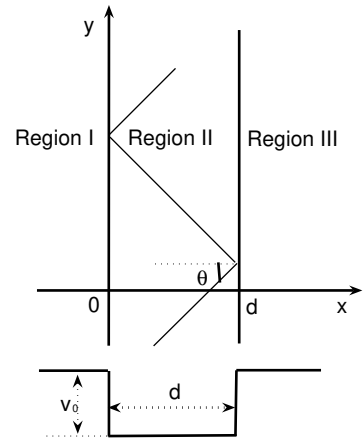}}
		\caption{\label{fig.1} Schematic diagram of the graphene waveguides.}
	\end{center}
\end{figure}

For simplicity, we consider a symmetric QW waveguide structure of
monolayer graphene, as shown in Fig. \ref{fig.1},
where the potential profile of the three zones are denoted by
\begin{equation}V(x)=
\begin{cases} V_0, & \text{otherwise}, \\0, &  0\leq x\leq d.
\end{cases}
\end{equation}
The electron wave with angle $\theta$ known as the $zig-zag$ angle
respect to the $x$ axis is incident to the QW, the direction of
guided mode propagation is $y$ axis, and there are two types of
situations: (a) when the zigzag angle is less than the total
internal reflection (TIR) angle, the modes become radiation modes;
(b) if the incident angle is more than the critical angle, there
will exist oscillating guided modes. What as follows we will focus
on the latter case. The TIR angle is defined by
$\sin\theta_c={\kappa_2}/{\kappa_1}$, where $\kappa_1=|E|/\hbar
v_F$, $\kappa_2=|E-V_0|/\hbar v_F$ are wave vectors in region II and
region III (or region I), E is the electron energy, $\hbar$ is
Planck's constant $h$ divided by $2\pi$, $v_F\approx10^6 m/s$ is
Fermi velocity. Carriers in graphene exhibit ballistic transport on
the submicron scale, and are governed by the following Dirac-type
equation:
\begin{equation}
 H=-i\hbar v_F \sigma\nabla+V(x),
\end{equation}
where $\sigma=(\sigma_x,\sigma_y)$ are the Pauli's matrices. The
Hamiltonian acts on the states expressed by the two-component
spinors $\Psi=[\Psi_A,\Psi_B]^T$, where $\Psi_A,\Psi_B$ represent
the smooth enveloping functions at the respective sublattice sites
of graphene. Due to translation invariance in the y direction, we
give the solution in the time-independent form
$\Psi_m(x,y)=\psi_m(x)e^{i\kappa_y y}$, $m=A,B$, thus obtain
\begin{eqnarray}
\begin{split}
-i\hbar
v_F(\frac{d\psi_B}{dx}+\kappa_y\psi_B)=(E-V(x))\psi_A, \\
-i\hbar v_F(\frac{d\psi_A}{dx}-\kappa_y\psi_A)=(E-V(x))\psi_B .
\end{split}
\end{eqnarray}
The character of the spinors solution in the three regions can be
written as:
\begin{equation}
\begin{split}
\psi_A(x)&=
\begin{cases} A e^{\alpha x}, & x<0, \\B \cos(\kappa_x x)+C\sin(\kappa_x x), &
0<x<d, \\De^{-\alpha (x-d)}, & x>d,
\end{cases}\\
\psi_B(x)&=
\begin{cases} -i A s'\frac{\delta-sin\theta}{sin\theta_c}e^{\alpha
x}, & x<0, \\is( B \sin(\kappa_x x+\theta)-C\cos(\kappa_x
x+\theta)), & 0<x<d,
\\is'D\frac{\delta+sin\theta}{sin\theta_c}e^{-\alpha (x-d)}, & x>d.
\end{cases}
\end{split}
\end{equation}
We define $s=sgn(E), s'=sgn(E-V_0),
 \delta=\sqrt{\sin^2\theta-\sin^2\theta_c}, \kappa_x=\kappa_1
\cos\theta, \kappa_y=\kappa_1 \sin\theta $, and $\alpha
=\sqrt{\kappa_1^2\sin^2\theta-\kappa_2^2}$ is the evanescent
coefficient when condition $0<V_0<2E$ is satisfied. Applying the
continuity of wave function at the interfaces $x=0$ and $x=d$, we
obtain the corresponding dispersion equation as follows,
\begin{equation}
\label{dispersion equation} \tan(\kappa_x
d)=\frac{ss'\cdot\kappa_x\sqrt{\kappa_y^2-\kappa_2^2}}{\kappa_1
\kappa_2-ss'\kappa_y^2},
\end{equation}
In order to describe the right-hand-side of Eq. (\ref{dispersion
equation}) as a function of $\kappa_x d$ and reveal the guided
modes, we make Eq. (\ref{dispersion equation}) in dimensionless form
\begin{equation}
F(\kappa_x d)=\frac{ss'\cdot(\kappa_xd)
\sqrt{(\kappa_1d)^2-(\kappa_xd)^2-(\kappa_2d)^2}}
{-ss'[(\kappa_1d)^2-(\kappa_xd )^2]+(\kappa_1d)\times(\kappa_2d) }.
\end{equation}

\begin{figure}[]
	\begin{center}
		\scalebox{0.50}[0.50]{\includegraphics{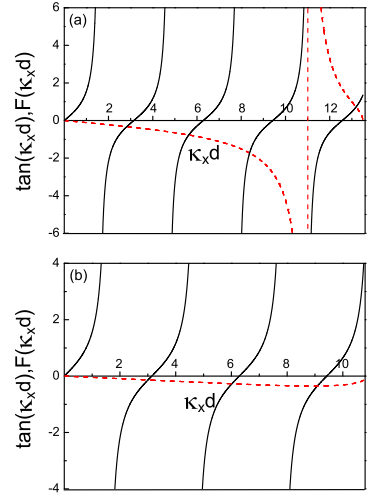}}
		\caption{\label{fig.2} Graphical determination of $\kappa_xd$ for oscillating
			guided modes, the solid and dashed curves correspond to the
			$\tan(\kappa_xd)$, $F(\kappa_xd)$ respectively, where physical
			parameters are chosen to be $U_0=50meV$, (a) ${\kappa_1 d}=4.96\pi$,
			${\kappa_2 d}=2.48\pi$, $ss'=1$, and (b) ${\kappa_1 d}=4\pi$,
			${\kappa_2 d}=2\pi$, $ss'=-1$.}
	\end{center}
\end{figure}

\begin{figure}[]
	\begin{center}
		\scalebox{0.42}[0.42]{\includegraphics{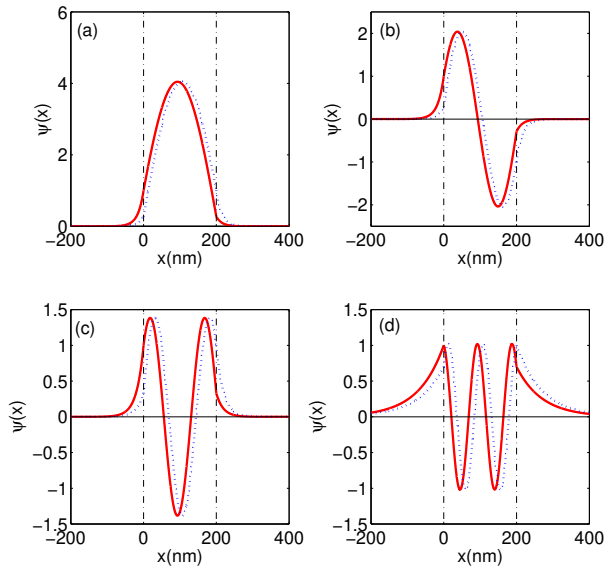}}
		\caption{\label{fig.3} The wave function of guided modes as function of distance of
			graphene waveguide corresponding to the intersections in Fig.
			\ref{fig.2}(a). The solid curve and the dashed curve correspond to
			$\psi_A$ and $-i\psi_B$, respectively. The physical parameters are
			$d=200nm$, (a) $ \kappa_xd=2.82$, $\theta=79.6^\circ$, (b)
			$\kappa_xd=5.63$, $\theta=68.8^\circ$, (c) $\kappa_xd=8.38$,
			$\theta=57.48^\circ$, (d) $\kappa_xd=13.2$, and $\theta=32.2^\circ$.}
	\end{center}
\end{figure}

The dispersion equation (\ref{dispersion equation}) is a
transcendental one and cannot be solved analytically, so we propose
a graphical method to determine the solution of $\kappa_x d$ for the
guided modes. Due to the different $ss'$, we will discuss the
properties of the guided modes in the following two cases,
respectively.

i) When $ss'=1 $, we get one of the dispersion equations. The
electron makes an intraband tunneling from an electron-like to
another electron-like, which corresponds to the case of classical
motion. To determine the guided modes, we plot the dependencies of
$F(\kappa_x d)$ and $\tan(\kappa_x d)$ on $\kappa_x d$ in Fig. 2(a).
The intersections show the existence of the guided modes. For the
given parameters, we can get four $\kappa_xd$ from the graph
directly. The profiles of waves function corresponding to the four
guided modes are shown in Fig. \ref{fig.3}.

\begin{figure}[]
	\begin{center}
		\scalebox{0.45}[0.45]{\includegraphics{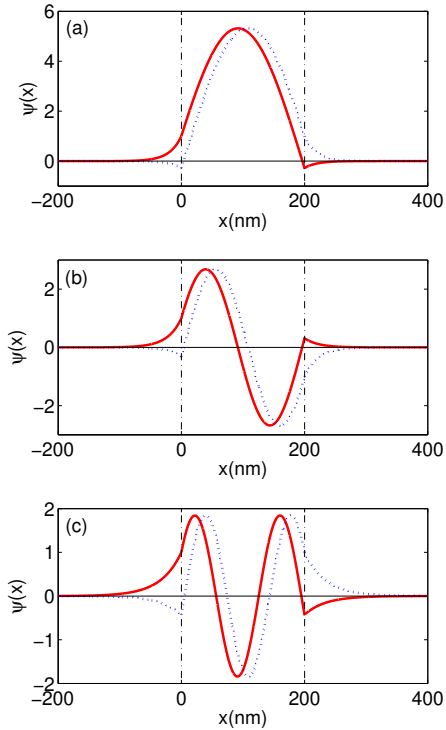}}
		\caption{\label{fig.4} The wave function of guided modes as function of distance of
			graphene waveguide corresponding to the intersections in Fig.
			\ref{fig.2} (b). The solid curve and the dashed curve correspond to
			$\psi_A$ and $-i\psi_B$, respectively. The physical parameters are
			$d=200nm$, (a) $\kappa_xd=3.0$, $\theta=76.2^{\circ}$, (b)
			$\kappa_xd=6.02$, $\theta=61.4^{\circ}$, (c) $\kappa_xd=9.08$,
			$\theta=43.7^{\circ}$.
		}
	\end{center}
\end{figure}

As shown in Fig. \ref{fig.3}, the states of $\psi_A$ and $-i\psi_B$
have similar characteristic, we discuss only the wave function
$\psi_A$. It is clearly seen that the electron waveguide supports
fundamental mode, first-order mode, second-order mode and
fourth-order mode, but the third-order mode is missing. In fact, the
hierarchy of the guided modes is dependent on the incident energy.
For a given QW electron waveguide, when the incident energy is not
sufficiently large with respect to the critical angle for the
third-order mode, this mode can not exist. It is seen that for
greater incident energy, the third-order mode will also be vanished.
In a word, the properties of the guided modes in the classical
motion case are analogous to those in the conventional QW electron
waveguide \cite{T.K2}. In addition, the tiny shift between the two
wave function is also demonstrated, because the electron velocity of
$\psi_A$ and $-i\psi_B$ are different.

ii) If $ss'=-1$, another dispersion equation is obtained. The
electron makes an intraband tunneling from an electron-like to
hole-like, which corresponds to the case of Klein tunneling. In this
case, the incident particles are predicted to tunnel through the
symmetric potential barriers with unit probability when the incident
energy $E<V_0$ is satisfied \cite{Na Ph}. This manifests the
combination of the conservation of the transverse momentum and the
absence of backscattering. Similarly the intersections of dashed
curves with solid curves indicate the existence of solutions for
guided modes, as shown in  Fig. \ref{fig.2} (b). Moreover, we could
get the $\kappa_xd$ from the graph directly, and describe the wave
function of the corresponding intersection in Fig. \ref{fig.4}.

From Fig. \ref{fig.4} it is demonstrated that the number of the
nodes of the guided modes $\psi_A$  begins from one and is
order-producing with the decreasing angle. In the case of Klein
tunneling, the properties of the QW electron waveguides are found to
be quite different. The conventional hierarchy of guided modes
disappears. Specifically, the fundamental node-less mode does not
exist at all, which is the particular property corresponding to the
Klein tunneling in graphene QW waveguide, since the fundamental mode
exists only when $E>V_0$ in the conventional QW electron waveguide.
Fig. \ref{fig.4} manifests an obvious shift between the wave
function $\psi_A$ and $-i\psi_B$, which is due to the different
speed between the electron for $\psi_A$ and that of the hole for
$-i\psi_B$. The wave function $\psi_A$ shows a peak closing to the
right interface between region II and region III , while the
$-i\psi_B$ shows a peak closing to the left interface between region
I and region II.

Finally, we should mention that when the graphene ribbons are narrow
with the widths in the range of micrometer or submicrometer, the two
types of edge states, which are current-carrying and dispersionless
surface states \cite{195408}, have strong influence on the transport
of the quantum Hall region. However, it is shown in this paper that
the oscillating guided modes guided by the electron wave have the
strong energy flux localized inside the well region of graphene
waveguides. In addition, the amplitude of the electron wave is
oscillating in the well region and is evanescent out of the well.
Thus, the influence of the edge states can be neglected in this
problem.

In conclusion, we have investigated the guided modes in graphene
waveguides. The oscillating guided modes of electron waves have
features analogous to those of classical guided modes in ordinary
dielectric structures. It is shown that the amplitude of electron
wave function is oscillating in the well region and is evanescent
out of the well. We obtain two different dispersion relations. In
the classical case the absence of third-order mode is demonstrated,
while for the Klein tunneling case, the fundamental mode is absent.
With the development of graphene, we hope that based on these
electron wave propagation characteristics, a wide variety of
graphene-based devices can be motivated by the application in the
control and guiding of ballistic electrons in the future.

This work was supported in part by the National Natural Science
Foundation of China (60806041, 60877055), the Shanghai Rising-Star
Program (08QA14030), the Science and Technology Commission of
Shanghai Municipal (08JC14097), the Shanghai Educational Development
Foundation (2007CG52), and the Shanghai Leading Academic Discipline
Program (S30105). X. Chen is also supported by Juan de la Cierva
Programme of Spanish Ministry of Science and Innovation.

\end{document}